# A Novel Technique to Observe Rapidly Pulsating Objects Using Spectral Wave-Interaction Effects

Ermanno F. Borra,
Centre d'Optique, Photonique et Laser, and Centre de Recherche en
Astrophysique du Québec
Département de Physique, Université Laval, Québec, Qc, Canada G1K 7P4
(email: borra@phy.ulaval.ca)

**ABSTRACT**

Conventional techniques that measure rapid time variations are inefficient or inadequate to discover and observe rapidly pulsating astronomical sources. It is therefore conceivable that there exist some classes of objects pulsating with extremely short periods that have not yet been discovered.

This article starts from the fact that rapid flux variations generate a spectral modulation that can be detected in the beat spectrum of the output current fluctuations of a quadratic detector. The telescope could observe at any frequency, although shorter frequencies would have the advantage of lower photon noise. The techniques would allow us to find and observe extremely fast time variations, opening up a new time window in Astronomy.

The current fluctuation technique, like intensity interferometers, uses second-order correlation effects and fits into the current renewal of interest in intensity interferometry. An interesting aspect it shares with intensity interferometry is that it can use inexpensive large telescope that have low-quality mirrors, like Cherenkov telescopes. It has other advantages over conventional techniques that measure time variations, foremost of which is its simplicity. Consequently, it could be used for extended monitoring of astronomical sources, something that is difficult to do with conventional telescopes.

Arguably, the most interesting scientific justification for the technique comes from Serendipity.

## 1. INTRODUCTION

While most physical domains (e.g. energy and angular resolution) have been deeply explored in Astronomy, the time domain is the least explored of the physical domains. Recognizing this, there currently are efforts made to explore further the short time domain (Barbieri et al, 2007, Deil et al. 2009).

The physical parameters of a large number of astronomical objects vary in time. Presently variable objects are detected by observing how intensity varies as a function of time. This works well for sources that have sufficiently slow time variations. However, present techniques become increasingly inefficient and eventually inadequate as the time scales of variation decrease. For example, as time scales decrease time sampling frequency must increase, giving an increasingly large quantity of data that can become difficult to store and analyze (Siemion et al. 2009). The next generation of astronomical instruments capable of high time resolutions is vey complex and will require very large telescopes (Barbieri et al. 2007). Detectors and electronics also have fundamental limitations because they cannot detect time variations below a certain value.

Because present techniques are inadequate to discover and observe extremely fast time variations, it is conceivable that there exist some classes of rapidly pulsating astronomical objects that have not yet been discovered. Also, some astronomical sources (e.g. millisecond pulsars) generate pulses that have complex structures within short time scales (nanoseconds). Because of the small time scales involved, it is difficult to observe these objects since observing time on a large telescope is required.

This article proposes a technique that could be used to observe and discover new classes of rapidly varying astronomical sources and find fine time structure in known ones (e.g. millisecond pulsars). It uses the fact that time variations of the intensity signal originating from a pulsating source modulate its spectrum. With the proposed technique, time variations would be detected in the spectrum of the fluctuations of the output current of a quadratic detector. A quadratic detector is an instrument that measures the square of the electromagnetic field amplitude averaged over a time considerably longer than the period of variation of the electromagnetic field. A photomultiplier is an example of quadratic detector used in the visible region of the electromagnetic spectrum. Because the technique uses second-order correlation spectroscopic effects it can use inexpensive telescopes having poor optical quality mirrors. It is possible to observe in any spectral region, from the ultraviolet to the radio region, although lower frequencies have an advantage.

## 2. BASIC THEORY

Consider a pulsating source having an electric field with a time dependent variation *E(t)* made of periodic pulses *V(t)* separated by a time interval $\tau$. The Array theorem of Fourier analysis gives us an equation to start from. The time variation *E(t)* is given by the convolution of *V(t)* with a comb function $\sum_m \delta(t-t_m)$ ) where $\delta(t-t_m)$ is the delta function and $t_m = (m-1)\tau$ and *m* is an integer number.

It can be shown (Borra 2010) that the Fourier transform of *E(t)* is then given by

$$H(\omega) = G(\omega)e^{i(N-1)\omega\tau/2}\left[\sin(\omega N\tau/2)/(\sin(\omega\tau/2))\right] \quad . \tag{1}$$

The signal *S($\omega$)* measured by astronomical quadratic detectors is proportional to *H($\omega$)H\*($\omega$)*, where the asterisk indicates complex conjugation. Using equation (1) one obtains

$$S(\omega) = S_1(\omega)\left[\sin(\omega N\tau/2)/(\sin(\omega\tau/2))\right]^2 \tag{2}$$

A figure in Borra (2010) shows the shapes of spectra generated by Equation (2) for several values of *N*. Equation (2) shows that *S($\omega$)* carries a signature that depends on the period $\tau$. The maxima of *S($\omega$)* are spaced by

$$\Delta\omega = 2\pi n/\tau, \tag{3}$$

where *n* is an integer number. Note that we are here discussing the same basic physics of the frequency comb technology, where a rapidly pulsed laser sends a sequence of very short optical pulses equally spaced in time which generates an optical spectrum made of a series of Dirac delta functions. Steinmetz et al. (2008) discuss astronomical applications of frequency combs.

Because the actual signal measured depends on the equipment used, we shall carry out a general discussion in terms of $H(\omega)$ only. A detailed discussion will be done below, starting from equation (1), only for the output current $I(t)$ of a quadratic detector that leads to detection of the spectral modulation of the current fluctuations. With increasing $N$ equation (1) predicts a shape having sharp peaks separated by $\Delta\omega = 2\pi n/\tau$. $S(\omega)$ and the absolute value of $H(\omega)$ closely resemble comb functions made of $\delta$ functions for $N>10$.

Time variations can be observed with spectroscopic equipment (e.g. a spectrograph) that detects a signal that depends on the spectral modulation of $S(\omega)$ (Borra 2010). However, because equation (3) shows that the frequency spacing $\Delta\omega$ is inversely proportional to the time between pulses $\tau$, it could only be directly observed for extremely short periods of the order of $10^{-12}$ seconds with standard spectroscopic equipment in the visual-infrared regions of the spectrum (Borra 2010). Although it could find new objects with extremely short period, it would be incapable of detecting values of $\tau$ between these extremely short values and the values of known rapidly pulsating sources (e.g. pulsars that have $\tau$ greater than 1 millisecond).

The detection technique that I propose in what follows allows detection of slower pulsators. It is based on a didactic experiment originally carried out by Alford & Gold (1958), hereafter referred to as AG, to measure the speed of light. Although the AG effect is unintuitive and seems to violate well-established physical principles, it is experimentally well-proven (Alford & Gold 1958, Basano & Ottonello 2000, section 3 in this article) and has firm theoretical foundations (Givens 1961, Mandel 1962). To understand the AG effect one must first accept the fact that *spectral* interference effects occur for optical path differences that are much larger than the temporal coherence length of the electromagnetic wave measured. Although this seems to be incompatible with the more familiar criterion of *classical* interferometry, it is proven by the cited experiments for *spectral* interferometry. Actually the optical path difference *must* be much larger than the temporal coherence length for *spectral* interference to occur (Mandel 1962). Secondly, one must understand that beams originating from sources that are incoherent according to classical definition *spectrally* interfere provided coherence is artificially provided by periodic pulsation (Givens 1961). Alford & Gold (1958) found interference between two optical independent beams originating from different sides of a spark, which would have been incoherent according to classical criteria. Coherence was artificially provided by pulsation. Our experiments in section 3 confirm this. Finally, one must understand that, because the electromagnetic waves in the primary spectrum beat among themselves, they generate a lower frequency beat spectrum. The spectral distribution of this beat spectrum can be obtained from intensity measurements with a square-law detector (e.g. a photomultiplier).

While Mandel (1962) gives a detailed theoretical justification of the AG effect, it is however mostly concerned with interfering beams from continuous (unpulsed) sources. Givens (1961) gives an analysis which assumes pulsed sources, is easier to follow, and gives a physical explanation that is adequate for our purpose. Therefore, to understand the principles of this non-intuitive technique let us follow Givens (1961) by considering the simple case where we only have two pulses. We assume classical wave interactions and the superposition principle that applies in classical electromagnetism. Using equation (1) with $N = 2$ and the trigonometric relation $sin(2\alpha) = 2 \, sin(\alpha)$ we obtain

$$H(\omega) = 2\cos(\omega\tau/2)G(\omega) \ . \tag{4}$$

Equation (4), like equation (1), predicts that the minima in the spectrum are separated by frequency intervals given by equation (3). For $\tau$ significantly larger than $10^{-12}$ seconds, the frequency separation becomes too small to be detected with classical optical spectroscopic techniques. However, we can borrow from Alford & Gold (1958) and measure the minima in the beat spectrum detected in the current fluctuations of a quadratic detector (a photomultiplier in their experimental setup). We would measure the current $I(t)$ and detect the spectral minima of the fluctuations of $I(t)$ that appear at much lower frequencies than those of $H(\omega)$.

Givens (1961) shows, starting from equation (4), that the frequency spectrum $I(\omega')$ of the output current $I(t)$ generated by the quadratic detector is given by

$$I(\omega') = (2\pi)^{1/2}\cos(\omega'\tau/2)\int a(\phi+\omega')a^*(\phi)G(\phi+\omega')G^*(\phi)d\phi, \tag{5}$$

where $a(\omega)$ is the bandpass of the detector, $\omega'$ is the beat frequency ($\omega >> \omega'$) and the asterisk indicates complex conjugation. Approximating $a(\omega)G(\omega)$ by a Gaussian having dispersion $\sigma$, we obtain

$$I(\omega') = K\cos(\omega'\tau/2)\exp[-\omega'^2/(2\sigma^2)], \tag{6}$$

where all the constants generated by the integration are grouped in the constant $K$. Equation (6) assumes that the two pulses have exactly the same shape and amplitude $V(t)$. However, this is not a stringent criterion (Alford & Gold 1958 and discussion below). The beat frequency $\omega'$ measured by the equipment placed after the detector, *is* given by the difference $\delta\omega$ between any two frequencies $\omega$ and $\omega + \delta\omega$ in the primary spectrum. The current $I(\omega')$ is spectrally modulated with maxima separated by $(2n+1)\pi/\tau$ (n is an integer number) and can be measured at a convenient low beat frequency $\omega'$ of our choice. Note that the beat frequency $\omega'$ can vary while the frequency of observation $\omega$ remains constant. Consequently, the spectral modulation can be measured at low frequencies $\omega'$ well outside the spectral bandpass of the light seen by the detector.

The above discussion applies to two pulses, while typical variable astronomical sources have periodic variations that are modeled by the Array theorem. The theory can readily be extended to consider periodic pulses having different amplitudes and shapes. To illustrate this, let us consider again the simple case where the 2 pulse separated by $\tau$. We have seen that it gave equation (4), thereby predicting a sinusoidal modulation. Another way of obtaining equation (6) is to repeat the analysis that leads to equation (1) for only 2 pulses. The sum of two pulses shifted by ± $\tau/2$ has a frequency spectrum given by the Fourier transform

$$H(\omega) = 1/(2\pi)^{1/2} \int_{-\infty}^{+\infty} [V(t+\tau/2) + V(t-\tau/2)] e^{-i\omega t} dt . \tag{7}$$

Applying the shift theorem to equation (7) gives

$$H(\omega) = (e^{i\omega\tau/2} + e^{-i\omega\tau/2}) G(\omega) = 2\cos(\omega\tau/2) G(\omega) , \tag{8}$$

which is the same equation as equation (4), also showing that the spectrum is modulated by a *cos( ωτ/2)* term. Let us now extend this analysis to the case where each of the two pulses is followed by an identical pulse displaced by $\delta\tau$. The Fourier transform now gives

$$H(\omega) = 1/(2\pi)^{1/2} \int_{-\infty}^{+\infty} [V(t+\tau/2) + V(t+\delta\tau+\tau/2) + V(t-\tau/2) + V(t+\delta\tau-\tau/2)] e^{-i\omega t} dt$$

(9)

Applying the shift theorem to equation (9) gives

$$H(\omega) = (e^{i\omega\tau/2} + e^{-i\omega\tau/2} + e^{i\omega\tau/2+\delta\tau} + e^{-i\omega\tau/2+\delta\tau}) G(\omega) \tag{10}$$

The result is a complex periodic shape that carries the signature of the time dependency of the pulses in equation (9). This can be seen by taking the real part of equation (10) and doing simulation with mathematical software. An analysis similar to the one leading to equation (6) would give a function *I(ω)* having a more complicated periodic structure than in equation (8). The experiments and presented in section 3 confirm that such a signal can be numerically predicted and detected in the spectrum of the current fluctuations.

Consider now two pulses having different shapes *V₁(t)* and *V₂(t)* . We can find a function *X(t)* such that

*V₂(t) = V₁(t ) + X(t).* (11)

Inserting *V₂(t)* in equation (7) we then obtain

$$H(\omega) = 1/(2\pi)^{1/2} \int_{-\infty}^{+\infty} [V_1(t+\tau/2) + V_1(t-\tau/2) + X(t-\tau/2)] e^{-i\omega t} dt . \tag{12}$$

This integral can be separated in the sum of two terms. The first one containing *V₁(t+ τ/2)* and *V₁(t- τ/2)* gives the same solution as in equation (8). The last term is the Fourier transform of *X(t)* which thus gives the spectral density of *X(t).* We therefore have

$$H(\omega) = (e^{i\omega\tau/2} + e^{-i\omega\tau/2})G(\omega) = 2\cos(\omega\tau/2)G(\omega) + 1/(2\pi)^{1/2} \int_{-\infty}^{+\infty} X(t-\tau/2)]e^{-i\omega t} dt \,.(13)$$

Equation (13) shows that the only effect is a decrease of the contrast of the spectral modulation by adding a background signal which is a function of $\omega$. The last term contains information about the spectral shape of the pulses. This theoretical analysis can be extended to more general cases but this is beyond the scope of this work.

### 3. EXPERIMENTS

Laboratory experiments have been carried out to demonstrate the technique. Fig. 1 shows the optical set up that was used to obtain the data. The light source is a diode laser operating at a wavelength of 1550 nm. The beam is pulsed using two independent commercial pulse generators generating pulses at a frequency of 10 KHz. The time delay between the 2 independent pulses can be adjusted at will. The light pulses are then sent into a Mach-Zehnder interferometer made of two optical fibers having respective lengths of 35 and 49 meters. This splits each pulse from the diode into two pulses separated by 70 nanoseconds. The setup therefore produces 4 pulses. Any of the pulses in the beam can be switched on and off with a Pockels cell. Finally, the signal is observed by a quadratic detector and the frequency spectrum of the output current fluctuations observed with a spectrum analyzer.

Fig. 2 shows two pairs of pulses, each generated by a different generator, separated by 300 ns. Each pair is split, by the interferometer, in two pulses separated by 70 ns. Fig. 3 shows the beat spectrum produced by the 4 pulses seen in Fig. 2. Fig. 4 shows the beat spectrum produced only by the first and last pulse seen in Fig. 2. The second and third pulses were cut off by the Pockels cell. We can see the type of sinusoidally modulated spectrum predicted by equation (5). The first and last pulse in Fig. 2 are separated by $\tau = 370$ nanoseconds, predicting a frequency separation of 2.70 MHz, in excellent agreement with the data that gives 2.83 MHz,

It is important to note that the first and fourth pulses shown in Fig. 2 are not coherent in the classical sense, since they are generated by two different electronic pulse generators. However, Fig. 4 shows interference patterns when the second and third pulses are cut off by the Pockels cell. Only the first and fourth pulses produce the spectral interference seen in these two figures. The data therefore confirm that spectral interference occurs among classically incoherent beams provided coherence is artificially induced by periodic pulsation, as stated in section 2. This is the same phenomenon that created interference between two optically independent beams originating from different sides of a spark in the experiment of Alford & Gold (1958). Coherence was also artificially provided by periodic pulsation in Alford & Gold (1958).

### 4. COMPARISON WITH OTHER TECHNIQUES

As with any novel technique, two questions now arise. Firstly, what are its advantages over techniques currently used that measure intensity variations as a function of time? Secondly, what are its inconveniences?

A major advantage of a spectral modulation technique comes from the fact that it would allow us to find objects that vary within short time scales that would be difficult or impossible to measure with current techniques. Observing directly the spectral modulation in the visible-infrared regions of the spectrum would allow us to observe periods as short as picoseconds and discover new classes of astronomical objects (Borra 2010). However, for substantially longer time separations, the minima would be too close to be separated with conventional spectrographs. The proposed technique that measures the spectrum of the output current fluctuations would be needed for significantly longer periods. Since there is no theoretical limitation to the frequency region that can be observed and thus one could observe anywhere from the X-rays to the radio region, the discussion below is of a general nature and makes no assumptions on the region of the electromagnetic spectrum that is observed.

Another advantage comes from the fact that it would be easier to implement than standard techniques because one could use simple and inexpensive equipment. Finding and observing fast pulsators requires very fast time sampling. This generates a large quantity of data that must be stored and analyzed. It requires sophisticated electronics, data storage and computing facilities (Siemion et al. 2009). The current fluctuation technique generates a far lower quantity of data, since it measures the low frequency spectrum of the current fluctuations. The data can be analyzed with simple equipment. For example, Alford & Gold (1958) detected the tightly separated pulses (time delays of $10^{-7}$ seconds) with the technology available in the middle of the last century. They used a tunable electric filter and sampled the signal at a few frequencies. The data discussed in section 3 of this article that detects pulses separated by times of the order of hundreds of nanoseconds was analyzed with a commercial spectrum analyzer.

Mandel (1962) shows that the AG effect uses second order correlation effects like intensity interferometry (Hanbury Brown 1968) and intensity correlation spectrometry (Goldberger, Lewis & Watson 1966). These are also referred to as two-photon experiments. Dravins (2008) gives a convenient summary of the potential future application of these techniques with the next-generation of very large astronomical telescopes. Like with intensity interferometry, this gives the proposed technique two major advantages. The first is that it is possible to observe inside a large bandpass, which helps the signal-to-noise ratio. The second, and more interesting, aspect comes from the fact that the requirements of the optical quality of the primary mirrors used are vastly relaxed so that large and inexpensive poor-quality reflectors can be used. One could use large inexpensive low surface quality specialized telescopes to gather large fluxes. In the optical and infrared region, one could use large inexpensive reflectors like those used in air Cherenkov telescopes currently used in gamma-ray astronomy. LeBohec & Holder (2006) have suggested using Aamospheric Cherenkov telescope arrays for optical intensity interferometry.

However, poor optical quality mirrors have drawbacks. The poor optical qualities of Cherenkov Telescopes spread out a stellar PSF over typically 3 arcminutes (Le Bohec & Holder 2006). This introduces a large contribution from the sky background that increases shot noise that thus decreases the limiting magnitude. Considering that at the zenith, in a moonless night, the night sky background in the V band is V = 22.5 per square arcseconds (Zombeck 1982), we find that the sky will contribution will be V= 11.5 within a 3 arcminutes diameter surface. Consequently, in principle, it would be difficult to go substantially fainter than V= 11.5 in the V band with a Cherenkov telescope. However, the issue of signal to noise ratio for second-order correlation techniques is quite complex. In particular, it depends on the degeneracy parameter, which is higher for non-thermal sources than thermal ones. This is discussed in section 5

Another effect one may worry about comes from the fact that some Cherenkov telescopes use spherical primary mirrors. Because the light reaches the focal position at different times, depending on which part of the primary mirror it has been reflected from, this introduces time dispersion. Typical time spreads in existing Cherenkov telescopes are of the order of nanoseconds (Akhperjanian & Sahakian 2004). For a parabolic multi-mirror telescope that uses multiple small spherical mirrors, the spherical shape of the individual mirror elements introduces some time dispersion of the order of a few tenths of nanoseconds. For example, the 17 m diameter f =1 MAGIC telescope has a time dispersion of 0.3 ns (Akhperjanian & Sahakian 2004). One may worry that this limits their performance in the search for nanosecond variability.

However, the theoretical analyses carried out by Givens (1961) and Mandel (1962) as well as section 2 show that the second-order effect discussed in this article is caused by the spectral modulation in the primary spectrum (equation 2) generated by first-order interference effects. This first-order spectral modulation is caused by the fact that power is transmitted to the detector at interfering frequencies that have phase differences that are either zero or even multiples of $\pi$, while the power is sent in a direction that does not reach the detector at frequencies that are out of phase by odd multiples of $\pi$. Note that this first-order interference effect is specifically invoked by Givens (1961). In a classical Michelson interferometer power is sent back to the source at frequencies that are out of phase by odd multiples of $\pi$. Because astronomical sources are extremely distant, the interfering beams are superposed like in a Michelson experiment, hence they arrive spectrally modulated prior to hitting the primary mirror of a telescope. The elongation of the pulses caused by the mirror is a linear effect: consequently, although the pulses are stretched in time, the spectral distribution (equation 2) is unchanged. Only non-linear effects can change the original spectral distribution. The effect of the spherical mirror can readily be modeled by breaking it into several zones (areas). Some mirrors (e.g., the MAGIC telescope) are actually made of several individual mirrors. One can then consider that every one of these subsections contributes individual sequences of pulses having the same time separations $\tau$ and the same spectral distributions given by equation (2). Because one then simply adds the same spectral distributions with the same separation, there will not be any degradation of the signal. Note that in the worse-case scenario that this theoretical interpretation is wrong, one would still be able to measure periods lower than ten nanoseconds.

The major inconvenience of the technique is that it uses wave-interaction effects like intensity interferometry. Mandel (1962) discusses the similarities between the AG effect and intensity interferometry and considers the effect of shot noise in spectral interferometry for continuous sources. Consequently, like intensity interferometry, photon shot-noise is a problem so that it requires very large light collectors. This inconvenience of intensity interferometry has been discussed at length by Hanbury Brown & Twiss (1958) who give the general formulae for the signal-to-noise ratio. However, this detailed analysis is cumbersome. Hanbury Brown (1968) gives a more succinct and easy to follow discussion. The next section discusses noise effects and the relation between the current fluctuation techniques and intensity interferometry.

## 5. NOISE AND SPATIAL COHERENCE EFFECTS

The theory in section 2 only considers wave interaction effects and neglects the effects of noise, in particular photon shot noise. The analysis of shot-noise effects in intensity correlation is very complicated. This can be appreciated by considering chapter 6 on problems involving high-order coherence and chapter 9 on the fundamental limits in the photoelectric detection of light in Goodman (1985) A detailed analysis of shot-noise effects in spectrum interferometry depends on the instruments and techniques used as well as on the physics responsible for the pulses (e.g. thermal or non-thermal sources) and is beyond the scope of this work, whose aim is simply to suggest the technique. We shall therefore confine ourselves to a brief generic discussion.

The effect of shot noise in spectral interferometry for continuous sources has been discussed by Mandel (1962) for the AG effect and Borra (1997, 2008) for continuous astronomical sources observed through a gravitational lens. These discussions are centered on considerations of the ratio R between the wave-interaction and shot noise terms of the spectral densities. The ratio R is given by (Mandel 1962, Purcell 1956, Borra 1997, Borra 2008)

$$R \leq \delta = \alpha \bar{I} 2\pi / \Delta\omega \quad , \tag{14}$$

where $\delta$ is the degeneracy parameter (Mandel 1962), $\alpha \bar{I}$ is the average count rate, $\Delta\omega$ the bandpass of observation and $2\pi/\Delta\omega$ gives the usual estimate of the coherence time approximated by the inverse of the width of the spectral bandpass. Equation (14) shows that the parameter $\delta$ is given by the average photon count rates inside the bandpass $\Delta\omega$ and that photon shot-noise dominates for $\delta<<1$. Shot-noise therefore overwhelms the signal when the counting rate is significantly below one count per coherence time interval $2\pi/\Delta\omega$, while the spectral modulation becomes easier to detect if $\delta>1$. This limit is similar to the one that applies to stellar intensity interferometry (Mandel 1962) and could be an inconvenience for observations of thermal sources where $\delta <<1$. However, the discussion below argues that the problem is less severe for pulsating sources and non-thermal sources.

Mandel (1962) justifies the use of equation (14) after a generalized abstract mathematical analysis that assumes the interference of two continuous beams originating from a thermal source observed through a Young interferometer. The intensity correlation effects considered by Mandel (1962) are thus due to interactions between intensity fluctuations, generated by wave interactions, which are correlated between the two beams. In a thermal source this effect is overwhelmed by shot noise. For example, in the visible spectral region a thermal source having the surface temperature of the sun has a $\delta$ of the order of $10^{-2}$. One can appreciate the effect of shot noise by considering equation (14) that shows that the ratio $R$ between the wave-interaction and shot noise terms is less than $\delta$. Long integration times and large telescopes are needed to obtain reasonable signal to noise ratios. In our case however we are dealing with pulsed sources which are likely to be generated by esoteric non-thermal processes. In a non-thermal mechanism (e.g. a laser-like mechanism) the value of $\delta$ is likely to be larger than the thermal value of $\delta$, which is a worse-case scenario, giving far better signal-to-noise ratio. For example, $\delta$ can be as high as $10^8$ for an optical laser (Mandel 1962).

Pulsation also helps by increasing the contrast of the signal, as shown by experiments of the AG effect made by Basano & Ottonello (2000) who found that the minima of the spectral modulation were sharper for a pulsed source than a continuous one. This is also shown by the fact that Alford & Gold (1958) easily detect the spectral modulation generated by a pulsating thermal source (an electric spark in their experiment). The frequency spectra in Alford & Gold (1958) show a strong modulation with deep minima.

Because the signal-to-noise ratio as a function of flux is a very complicated issue, which depends on the degeneracy parameter $\delta$ as well as photon shot-noise from the background, it is difficult to give limiting fluxes. This can be appreciated by considering that the ratio $R$ between the wave-interaction and shot noise (equation 14) depends on the parameter $\delta$ that can be $<<1$ for thermal sources and as high as $10^8$ for lasers. Consider also that a background shot noise can come from the sky as well as from an astronomical object (e.g. the nucleus of a galaxy) in which the pulsator resides. We can however obtain order of magnitude estimates by considering the estimates in Borra (1997) who assumes non-pulsating thermal sources that interfere. He finds that a 100-m radio telescope observing at 1 GHz, would be able to detect the spectral modulation to 4 milliJy. Because flux scales as the square of the diameter of the mirror and $\delta$ is inversely proportional to the frequency, one can readily extrapolate to different frequencies and diameters. For example, we find that a 30-m diameter telescope observing at a wavelength of 1 mm would be able to observe to 1 Jy. Note, however that this estimate for a non-pulsating thermal source does not take into account the photon shot noise from the background, which decreases the signal to nose ratio and thus increases the limiting flux. On the other hand, a pulsating non-thermal source would be detectable to fainter fluxes.

The theoretical analysis in section 2 does not take spatial coherence effects into account. Mandel (1962) considers the effect of spatial coherence, showing that they are strongly dependent on the extendedness of the source. However, his discussion assumes continuous interfering beams, while in our case we have interference of pulsed sources which allows interference of classically incoherent sources Furthermore, Mandel's abstract analysis does not specify the frequency at which the spatial correlation coefficient must be evaluated. Borra (2008) argues that the spatial coherence coefficient should depend on the beat frequency $\omega'$ and not the frequency of observations $\omega$. This renders spatial coherence effects significantly less important (Borra 2008).

Our discussion ignores the effects of the interstellar and interplanetary media and of the Earth atmosphere, which introduce scintillation that can affect the signal. Details of these effects depend on the frequency of observations and are beyond the scope of this work

## 6. ASTRONOMICAL APPLICATIONS

An important question that arises concerns the usefulness of the technique in Astronomy. At this point in time, millisecond pulsars are the astronomical pulsating objects with the shortest periods. The proposed technique does not offer significant advantage at these time scales. It could however be used to find and study structure in very short time scales in known pulsating sources such as pulsars, as discussed in section 2. In that case, the technique has, over intensity measurements, the advantage of simplicity mentioned in section 4. The pulses seen in pulsars have a complicate time structure. For example, echoes have been seen to follow the main pulses (Lyne,. Pritchard, Graham-Smith. 2001). The extreme nanopulses detected by Hankins et al. (2003) are unresolved at their 0.4 ns resolution Finer structure could be found and studied. The data presented in section 3 detect structures with separations less than hundred nanoseconds. The 4 pulses that make the structures are produced at a frequency of 10 KHz. This shows that fine time structure could be found in pulsars (or faster pulsators) with this technique.

For a pulsating source with a varying period, pulsation would be detected, provided the variation is sufficiently small. The technique would however be incapable of detecting rapid phenomena that are totally random.

Although the technique may be used at any frequency, lower frequencies have the advantage that the degeneracy parameter (equation 14) increases with decreasing frequency. Observations in the nanosecond range have been already carried out with state-of-the art equipment in the radio region (Hankins, Kern. Weatherall & Eilek 2003) so that one may question its usefulness in the radio region. An advantage comes from the fact that one could use simple and inexpensive instrumentation (e.g. commercially available correlators). Certainly, the technique would be advantageous in the infrared to millimeter bands where one could search for objects undergoing very rapid time variations.

Because there are no known pulsating sources with time scales shorter than milliseconds, serendipity is, at this point in time, the most interesting scientific justification for a search using the technique. A recent paper by Fabian (2010) discusses at length the role that serendipity has historically played in Astronomy, showing that chance discoveries play an important role. He notes that the time domain is the least explored one in astronomy and continues to be rich in discoveries.

Although, arguably, serendipity is, at this point in time, the most scientifically interesting justification for a search, it must make a minimum of physical sense; while at the same time one must not be too narrow-minded. For example, let us not forget that astronomical phenomena with periods of fractions of seconds were not known before 1950 and the discovery of Pulsars was totally unexpected. A potential problem comes from the fact that short time scales imply small regions and therefore high energy densities. If such a small pulsating source is a black-body emitting in the optical part of the spectrum (i.e., with a temperature of the order of thousands of K) it shall be exceedingly faint because of its small size. While there clearly is a major problem for thermal sources, it is less of a problem for non-thermal sources. Indeed, non-thermal sources with implied brightness temperatures of 2 $\times 10^{41}$ K have been observed in nanosecond-resolution observations of pulsars at radio frequencies (Hankins et al. 2003). Borra (2010) discusses the energy density issue in greater details. If such a non-thermal mechanism exists in pulsars, why not in other yet-unknown pulsators?

## 7. CONCLUSION

The novelty of this work comes from the suggestion that very rapid periodic flux variations in astronomical sources could be detected in the spectrum of the output current fluctuations of a quadratic detector.

The current fluctuation technique has several advantages over techniques presently used. A major one comes from the fact that it uses simple and inexpensive equipment, making it easy to implement. Another one is that, like the intensity interferometer (Hanbury Brown 1968), it uses second order correlations. The requirements of the optical quality of the mirrors used are thus vastly relaxed so that large and inexpensive poor-quality reflectors. For example, Cherenkov telescopes used for $\gamma$–ray astronomy, could be used to observe in the optical or infrared regions of the spectrum. Because the technique uses simple inexpensive instrumentation and can use inexpensive poor-surface-quality mirrors, it could be used for extended monitoring of light sources, something that is hard to do with conventional expensive telescopes, where it is difficult to obtain observing time. Its major disadvantage, like in intensity interferometry, comes from photon shot noise. However, it should be less of a problem than in intensity interferometry because the sources of interest would likely to be non-thermal, a factor that increases the degeneracy factor (see section 5).

Although the technique can be used to find very fast periodic pulsators (e.g. periods shorter than nanoseconds), it actually is more versatile than that. Fine time structure in slowly pulsating sources (e.g. millisecond pulsars) could be found and studied. The data presented in section 3 detect structures with separations less than one hundred nanoseconds. The 4 pulses that make the structures are produced at a frequency of 10 KHz, showing that fine time structure could be found in pulsars (or faster pulsators) with this technique.

The most interesting aspect of the technique is that it opens up the possibility of finding new classes of objects undergoing extremely fast pulsations, ranging from fractions of nanoseconds to fractions of milliseconds. This would open up a new astronomical window: the short-time-variations window. In the history of Astronomy, whenever a new window was opened, unexpected discoveries were made. This is the case for example in Radio Astronomy where quasars and pulsars were discovered and for the gamma rays, where gamma ray bursters were discovered. We should expect the unexpected.

## ACKNOWLEDGEMENTS

This research has been supported by the Natural Sciences and Engineering Research Council of Canada. The data presented in section 3 were obtained by Jean-François Brière.

## REFERENCES

Akhperjanian, A. & Sahakian, V., 2004 , Astropart. Phys., 21, 149
Alford, W. P., Gold, A.,1958, Am.J.Phys., 26, 481
Basano, L., Ottonello, P., 2000, Am.J.Phys., 68, 325
Barbieri, C. et al., 2007, J. of Modern Optics, 54, 191
Hanbury Brown, R., 1968, ARA&A, 6, 13
Borra, E.F., 2010, A&A Letters, In Press
Borra, E.F., 1997, MNRAS, 289, 660
Borra, E.F. 2008, MNRAS, 389, 364
Deil, C., Domainko, W., Hermann,G., Clapson, A. C., Förster, A, van Eldik, C., Hofmann, W. 2009, Astropart. Phys. 31, 156
Dravins, D. 2008, in High Time Resolution Astrophysics, eds. D.Phelan, O. Ryan, & A. Shearer; Astrophys. Space Sci. Lib. 351, 95
Fabian, A, C. 2010, in: "Serendipity. The Darwin College Lectures", eds. M. de Rond & I. Morley, Cambridge University Press, New York
Givens., M. P.,1961, J.Opt.Soc.Am., 51, 1030
Goldberger, M.L., Lewis, H.W., & Watson, K.M., 1966, Phys. Rev. 142, 25
Goodman, J.W., 1985, Statistical Optics. John Wiley & Sons. New York.
Hanbury Brown, R., 1968, ARA&A, 6, 13
Hanbury Brown, R., Twiss, R. Q., , 1958, *Proc. Roy. Soc. A,* 248, 199
Hankins, T. H., Kern, J. S., Weatherall, J. C. & Eilek, J. A., 2003, Nature, 422, 141.
Lyne, A. G. , Pritchard, R. S., Graham-Smith, F, MNRAS, 2001, 321, 67
LeBohec S., Holder J., 2006, ApJ 649, 399.
Mandel L., 1962, J.Opt.Soc.Am., 52, 1335
Purcell, E.M., 1956, Nature, 178, 1449
Siemion, A. et al. 2009, arXiv:0811.3046, to appear in Acta Astronautica
Steinmetz, T., et al., 2008, Science 321,1335
Zombeck. M.V. Handbook of Space Astronomy & Astrophysics, 1982, Cambridge University Press, New York

FIGURES

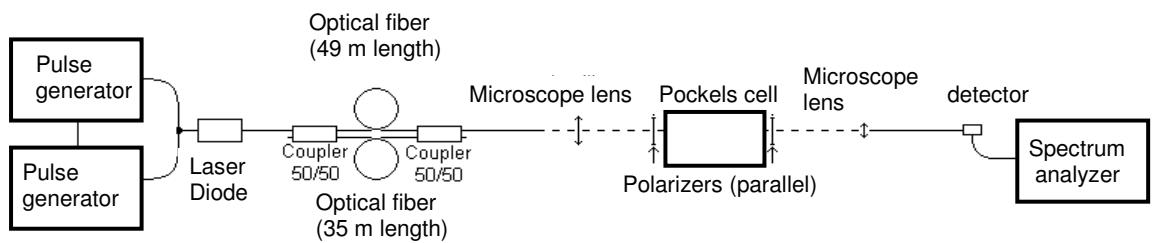

Figure 1 : Experimental setup used to produce 4 pulses. See text for more details

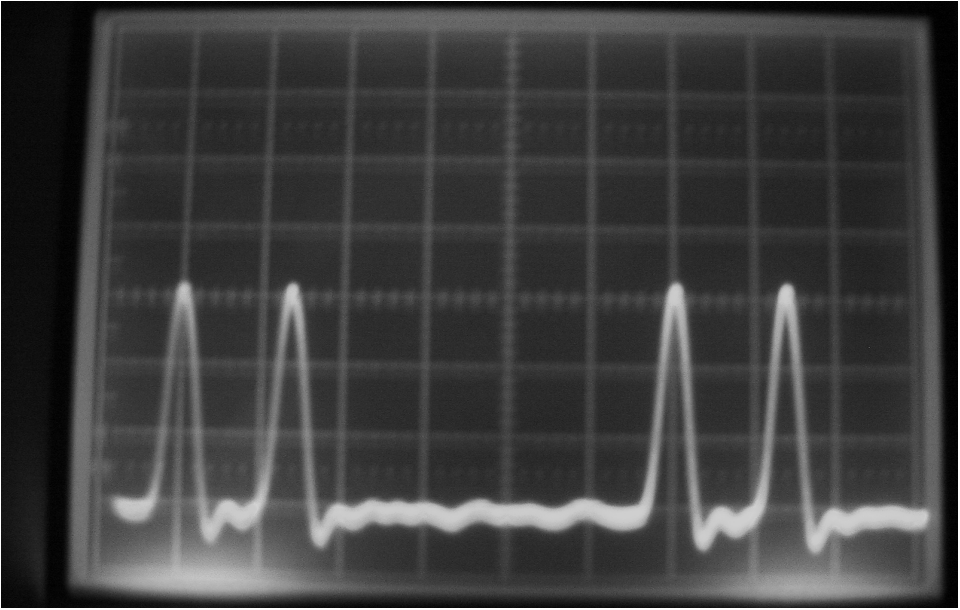

Figure 2 : It shows the pulses that generate the spectra displayed in figures 3 and 4. The pairs of pulses are separated by 300 ns and the members of a pair are separated by 70 ns. The horizontal scale is 50 nanoseconds/division.

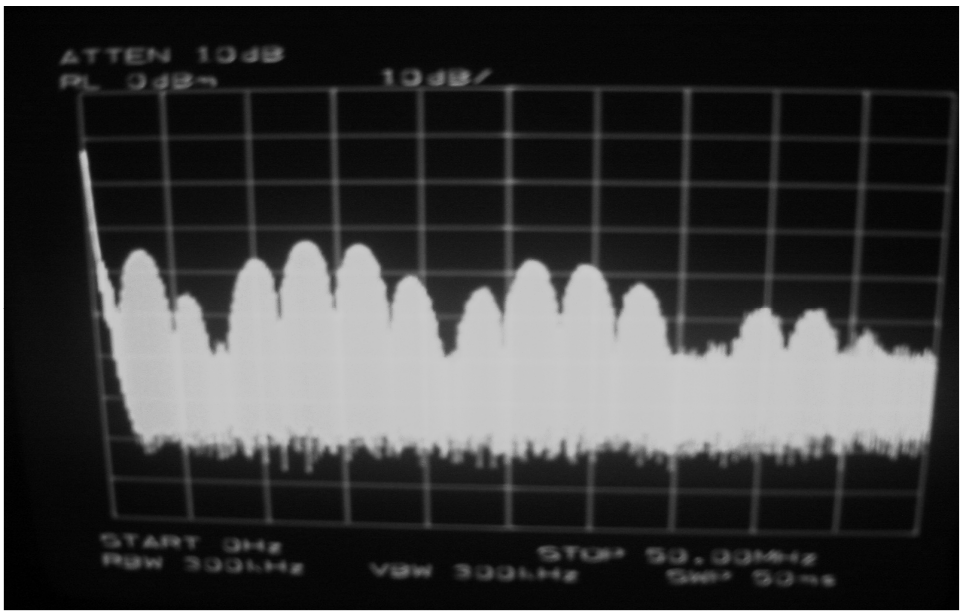

Figure 3. Beat spectrum produced by the pulses shown in figure 2. The horizontal scale varies from 0 MHz to 50 MHz. The vertical scale is logarithmic

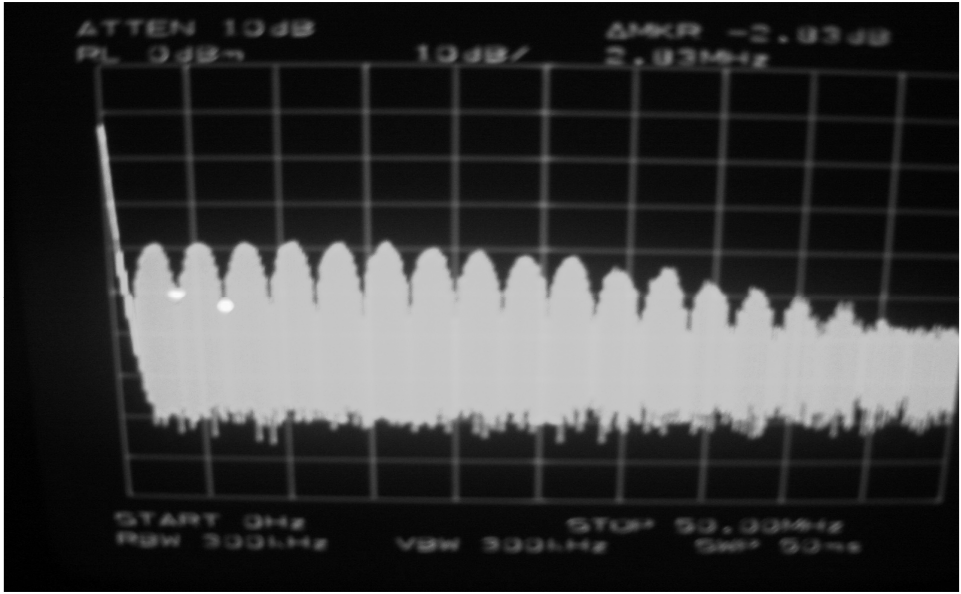

Figure 4 : Beat spectrum produced only by the first and last pulse shown in figure 2 which are separated by 370 ns. The second and third pulses have been cut by the Pockels cell. The horizontal scale varies from 0 MHz to 50 MHz. The vertical scale is logarithmic